\documentclass[a4paper,10pt,twocolumn]{article}
\usepackage{authblk}
\usepackage{graphicx}
\usepackage{latexsym}
\usepackage[fleqn]{amsmath}
\usepackage[psamsfonts]{amssymb}
\usepackage{float}
\usepackage[cm]{fullpage}
\usepackage{fancyvrb}
\fvset{fontsize=\small}

\title{[Technical Report]\\A Multiprecision Matrix Calculation Library and Its Extension Library for a Matrix-Product-State Simulation of Quantum Computing}
\author{Akira SaiToh}
\date{14 November 2011}
\affil{Research Center for Quantum Computing,
Interdisciplinary Graduate School of Science and Engineering, Kinki
University, 3-4-1 Kowakae, Higashi-Osaka, Osaka 577-8502, Japan}
\begin{document}
\twocolumn[
\begin{@twocolumnfalse}
\maketitle
\begin{abstract}
A C++ library, named ZKCM, has been developed for the purpose of multiprecision
matrix calculations, which is based on the GNU MP and MPFR libraries. It is
especially convenient for writing programs involving tensor-product operations,
tracing-out operations, and singular-value decompositions. Its extension
library, ZKCM\_QC, for simulating quantum computing has been developed
using the time-dependent matrix-product-state simulation method.
This report gives a brief introduction to the libraries with sample programs.

~~
\end{abstract}
\end{@twocolumnfalse}
]

\section{Introduction}
Accuracy of simulation is often of serious concern when small
differences in matrix elements result in physically important
phenomena of one's interest. There are several programming libraries, e.g.,
Refs.\ \cite{FMZM,CLN,Exflib,MPACK}, useful for high-precision computing
for this purpose. Among them, the library named ZKCM
library \cite{ZKCM}, which I have been developing, is a C++ library for
multiprecision complex-number matrix calculations. It provides
several functionalities including singular value decomposition,
tensor product calculation, and tracing-out
operations. It is based on the GNU MP (GMP) \cite{GMP} and MPFR
\cite{MPFR} libraries, which are commonly included in recent
distributions of UNIX-like systems.

There is an extension library named ZKCM\_QC. This library is 
designed for simulating quantum
computing \cite{Gruska,NC2000}. It uses a matrix product state (MPS)
\cite{WH93,V03} to represent a pure quantum state.
The MPS method is recently one of the standard methods for
simulation-physics software \cite{ALPS}. As for other methods effective
for simulating quantum computing, see, e.g., Refs.\ \cite{VMH03,AG04}.
With ZKCM\_QC, one may use quantum gates in $\rm{U(2)}$, ${\rm U(4)}$,
and ${\rm U(8)}$ as elementary gates. Indeed, in general, quantum
gates in ${\rm U(2)}$ and ${\rm U(4)}$ are enough for universal quantum
computing, but we regard quantum gates in ${\rm U(8)}$ also as
elementary gates so as to reduce computational overheads.

A simulation of quantum computing with MPS is known for its
computational efficiency in case the Schmidt ranks are kept small
during the simulation \cite{V03,KW04}.
Even for the case slightly large Schmidt ranks are involved, it is
not as expensive as a simple simulation. This is obvious from
the theory which is briefly explained in Sec.\ \ref{sectheory}.

This contribution is intended to provide a useful introduction
for programming with the libraries. Section\ \ref{secZKCM} describes
an example of simulating an NMR spectrum in a simple model using
the ZKCM library. Section\ \ref{secZKCMQC} shows an overview of the
theory of the MPS method and an example of simulating a simple quantum
circuit using the ZKCM\_QC library. Effectiveness of the use of the
libraries manifested by the examples are summarized in Sec.\ \ref{secSummary}.

\section{ZKCM Library}\label{secZKCM}
The ZKCM library is designed for general-purpose matrix calculations.
This section concentrates on its main library. It consists of two major
C++ classes: zkcm\_class and zkcm\_matrix. The former class is a class of
a complex number. Many operators like ``+='' and functions like
trigonometric functions are defined for the class. The latter class is a
class of a matrix. Standard operations and functions like matrix inversion
are defined. In addition, the singular-value decomposition of a
general matrix, the diagonalization of an Hermitian matrix, 
discrete Fourier transformation, etc., are defined for the class.
A detailed document is placed in the ``doc'' directory of the
package of ZKCM. We will next look at a simple example to
demonstrate the programming style using the library.

\subsection{Program example}
Here, a sample program ``NMR\_spectrum\_simulation.cpp''
found in the ``samples'' directory of the package of ZKCM
is explained. This program generates a simulated FID
spectrum of liquid-state NMR for the spin system consisting of a
proton spin with precession frequency $w_1=$~400 MHz (variable w1 in
the program) and a ${}^{13}{\rm C}$ spin with precession frequency $w_2=$~125 MHz
(variable w2) at room temperature (300 K) (variable T).
A J coupling constant $J_{12}=140$ kHz (variable J12) is considered
for the spins.

The first line of the program is to include a header file of ZKCM:
\begin{Verbatim}
#include "zkcm.hpp"
int main(int argc, char *argv[])
{
\end{Verbatim}
In the \verb|main| function, the internal precision
is set to 280 bits for floating-point computation by
\begin{Verbatim}
  zkcm_set_default_prec(280);
\end{Verbatim}
In the subsequent lines, Pauli matrices $I, X, Y, Z$
are generated. For example, $Y$ is generated as
\begin{Verbatim}
  zkcm_matrix Y(2,2);
  Y.set(zkcm_class(0.0,-1.0),0,1);
  Y.set(zkcm_class(0.0,1.0),1,0);
\end{Verbatim}
Similarly, the $Y_{90}$ pulse is generated as
\begin{Verbatim}
  Yhpi.set(sqrt(zkcm_class(0.5)),0,0);
  Yhpi.set(sqrt(zkcm_class(0.5)),0,1);
  Yhpi.set(-sqrt(zkcm_class(0.5)),1,0);
  Yhpi.set(sqrt(zkcm_class(0.5)),1,1);
\end{Verbatim}
Other matrices are generated by similar lines.
After this, values of constants and parameters are set.
For example, the Boltzmann constant $k_{\rm B}$ [J/K] is generated as
\begin{Verbatim}
  zkcm_class kB("1.3806504e-23");
\end{Verbatim}
(Several lines are omitted in this explanation.)
The Hamiltonian $H$ in the type of \verb|zkcm_matrix| is made as
\begin{Verbatim}
  H = w1 * tensorprod(Z,I) + w2 * tensorprod(I,Z)
      + J12 * tensorprod(Z,Z);
\end{Verbatim}
This is used to generate a thermal state $\rho$:
\begin{Verbatim}
  zkcm_matrix rho(4,4);
  rho = exp_H((-hplanck/kB/T) * H);
  rho /= trace(rho);
\end{Verbatim}
Here, \verb|exp_H| is a function to calculate the exponential of an
Hermitian matrix and \verb|hplanck| is the Planck constant
($6.62606896\times 10^{-34}$ Js).
The sampling time interval $dt$ to record the value of $<X>$ for the proton
spin is set to $0.145/w_1$ (any number sufficiently smaller than $1/2$
might be fine instead of $0.145$) by
\begin{Verbatim}
  zkcm_class dt(zkcm_class("0.145")/w1);
\end{Verbatim}
The number of data to record is then decided as
\begin{Verbatim}
  int N = UNP2(1.0/dt/J12);
\end{Verbatim}
Here, function UNP2 returns the integer upper nearest power of 2 for a
given number.
Now arrays to store data are prepared as row vectors.
\begin{Verbatim}
  zkcm_matrix array(1, N), array2(1,N);
\end{Verbatim}
The following lines prepare the X, Y, and $Y_{90}$-pulse operators
acting only on the proton spin.
\begin{Verbatim}
  zkcm_matrix X1(4,4), Y1(4,4), Yhpi1(4,4);
  X1 = tensorprod(X, I);
  Y1 = tensorprod(Y, I);
  Yhpi1 = tensorprod(Yhpi, I);
\end{Verbatim}
To get an FID data, we firstly tilt the proton spin by the ideal pulse.
\begin{Verbatim}
  rho = Yhpi1 * rho * adjoint(Yhpi1);
\end{Verbatim}
Now the data of time evolution of $<X>$ of the proton spin under the Hamiltonian
$H$ is recorded for the time duration ${\rm N} \times {\rm dt}$ using
\begin{Verbatim}
  array = rec_evol(rho, H, X1, dt, N);
\end{Verbatim}
We now use a zero-padding for this ``array'' so as to enhance the resolution.
This will extend the array by N zeros. 
\begin{Verbatim}
  array2 = zero_padding(array, 2*N);
\end{Verbatim}
To obtain a spectrum, the discrete Fourier transformation is applied.
\begin{Verbatim}
  array2 = abs(DFT(array2));
\end{Verbatim}
The ``array2'' is output to the file ``example\_zp.fid'' as an FID data
with $df = 1/(2\times 2N\times dt)$ as the frequency interval, in the
Gnuplot style by
\begin{Verbatim}
  GP_1D_print(array2, 1.0/dt/zkcm_class(2*N*2),
              1, "example_zp.fid");
\end{Verbatim}
At last, the function ``main'' ends with \verb|return 0;|.
The program is compiled and executed in a standard
way.\footnote{To make an executable file, the library flags typically
``-lzkcm -lm -lmpfr -lgmp -lgmpxx'' are probably required. As for
ZKCM\_QC, additionally ``-lzkcm\_qc'' should be specified.}

\begin{figure}[H]
\begin{center}
\includegraphics[width=79mm]{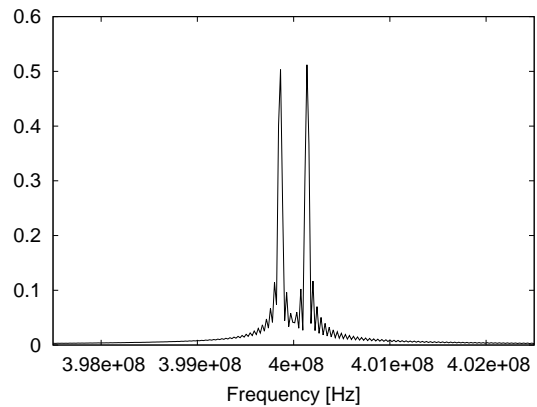}
\end{center}
\caption{\label{figzp} Plot of the simulation data stored in ``example\_zp.fid''.
See Software information for the environment where this simulation has been performed.}
\end{figure}

The result stored in ``example\_zp.fid'' is visualized by Gnuplot as
shown in Fig.\ \ref{figzp}.
It should be noted that we did not employ a high-temperature
approximation. Under a high-temperature approximation, the first order
deviation $-\beta H$ of $\exp(-\beta H)$ (here, $\beta=h/(k_{\rm B}T)$)
is considered as a deviation density matrix and calculations are
performed using the normalized deviation density matrix $-H/{\rm const.}$
This approximation is commonly used \cite{GAMMA} but it cannot be
used for simulations for low temperature.
An advantage of using ZKCM for simulating NMR spectra
is that the temperature can be chosen. This is possible because of
high accuracy in computing the exponential of a Hamiltonian.

\section{ZKCM\_QC library}\label{secZKCMQC}
The ZKCM\_QC library is an extension of the ZKCM library.
It has several classes to handle tensor data useful for
the time-dependent MPS simulation of a quantum circuit.
Among the classes, the ``mps'' class and the ``tensor2'' class
will be used by user-side programs. The former class
conceals the complicated MPS simulation process and enables 
writing programs in a simple manner. The latter class is
used to represent two-dimensional tensors which are often 
simply regarded as matrices. A quantum state during an MPS
simulation is obtained as a (reduced) density matrix in the
type of tensor2. For convenience, there is a function to
convert a matrix in the type of tensor2 to the type of zkcm\_matrix.

More details of the classes are explained in the document placed
at the ``doc'' directory of the ZKCM\_QC package.

We briefly overview the theory of the MPS simulation before
introducing a program example since understanding the behavior of a
library leads to a better programming in general.
\subsection{Brief overview of the theory of time-dependent MPS simulation}\label{sectheory}
Consider an $n$-qubit pure quantum state
\[
|\Psi\rangle=\sum_{i_0\cdots i_{n-1}=0\cdots0}^{1\cdots1}
c_{i_0\cdots i_{n-1}}|i_0\cdots i_{n-1}\rangle
\]
with $\sum_{i_0\cdots i_{n-1}}|c_{i_0\cdots i_{n-1}}|^2=1$.
If we keep this state as data as it is, updating the data for each
time of unitary time evolution spends $O(2^{2n})$ floating-point
operations. To avoid such an exhaustive calculation, in the
matrix-product-state method, the data is stored as a kind of compressed
data. The state can be represented in the form
\begin{equation}\label{eq1}
\begin{split}
 |\Psi\rangle&=\sum_{i_0=0}^1\cdots\sum_{i_{n-1}=0}^1\biggl[
\sum_{v_0=0}^{m_0-1}\sum_{v_1=0}^{m_1-1}\cdots\sum_{v_{n-2}=0}^{m_{n-2}-1}\\
&~~Q_0(i_0,v_0)V_0(v_0)Q_1(i_1,v_0,v_1)V_1(v_1)\cdots \\&~~\cdots 
Q_{s}(i_{s},v_{s-1},v_s)V_s(v_s)\cdots\\&~~\cdots V_{n-2}(v_{n-2})Q_{n-1}(i_{n-1},v_{n-2})\biggr]
|i_0\cdots i_{n-1}\rangle,
\end{split}\end{equation}
where we use tensors $\{Q_s\}_{s=0}^{n-1}$ with parameters $i_s, v_{s-1}, v_s$
($v_{-1}$ and $v_{n-1}$ are excluded) and $\{V_s\}_{s=0}^{n-2}$ with parameter
$v_s$; $m_s$ is a suitable number of values assigned to $v_s$
with which the state is represented precisely or well approximated.
This form is one of the forms of matrix product states (MPSs). The
data are compressed to tensor elements. We can see that neighboring
tensors are correlated to each other; the data compression is owing
to this structure.

Let us explain a little more details:
$Q_s(i_s,v_{s-1},v_{s})$ is a tensor with
$2\times m_{s-1} \times m_{s}$ elements; $V_s(v_s)$ is a tensor
in which the Schmidt coefficients for the splitting between the $s$th
site and the $(s+1)$th site ({\em i.e.}, the positive square roots of non-zero
eigenvalues of the reduced density operator of qubits $0,\ldots,s$) are
stored. This implies that, by using $V_s$ and eigenvectors
$|\Phi_{v_s}^{0\ldots s}\rangle$ ($|\Phi_{v_s}^{s+1\ldots n-1}\rangle$)
of the reduced density operator $\rho^{0\ldots s}$ ($\rho^{s+1\ldots n-1}$)
of qubits $0\,\ldots,s$ ($s+1,\ldots,n-1$), the state can also be written
in the form of Schmidt decomposition
\begin{equation}
 |\Psi\rangle = \sum_{v_s=0}^{m_s-1}V_s(v_s)
|\Phi_{v_s}^{0\ldots s}\rangle |\Phi_{v_s}^{s+1\ldots n-1}\rangle.
\end{equation}
In an MPS simulation, very small coefficients and corresponding
eigenvectors are truncated out unlike a usual Schmidt decomposition.

The advantage to use the MPS form is that we have only to handle
a small number of tensors when we simulate a time evolution under
a single quantum gate. For example, when we apply a unitary operation
$\in {\rm U}(4)$ acting on, say, qubits $s$ and $s+1$,
we have only to update the tensors $Q_s(i_s,v_{s-1},v_s)$, $V_s(v_s)$,
and $Q_{s+1}(i_{s+1},v_s,v_{s+1})$.
For the details of how tensors are updated, see Refs.\ \cite{V03,S06}.
The simulation of a single quantum gate $\in {\rm U}(4)$ spends
$O(m_{\rm max}^3)$ floating-point operations where $m_{\rm max}$
is the largest value of $m_s$ among the sites $s$.
(Usually, unitary operations $\in {\rm U}(2)$ and those $\in {\rm U}(4)$
are regarded as elementary quantum gates.)
A quantum circuit constructed by using at most $g$
single-qubit and/or two-qubit quantum gates can be simulated within the
cost of $O(g n m_{\rm max,max}^3)$ floating-point operations,
where $n$ is the number of wires and $m_{\rm max,max}$ is the largest
value of $m_{\rm max}$ over all time steps. 

The computational complexity may be slightly different for each software
using MPS. In the ZKCM\_QC library, we have functions to apply quantum gates
$\in {\rm U}(8)$ to three chosen qubits. Internally, three-qubit gates
are handled as elementary gates. This makes the complexity a little
larger. A simulation using the library spends $O(g n m_{\rm max,max}^4)$
floating-point operations, where $g$ is the number of single-qubit, two-qubit,
and/or three-qubit gates used for constructing a quantum circuit.

The MPS simulation process, which is in fact often complicated, can be
concealed by the use of ZKCM\_QC. One may write a program for quantum circuit
simulation in an intuitive manner. Here is a very simple example.
\subsection{Program example}
The following program is placed at the ``samples'' directory
of the ZKCM\_QC package. It utilizes several matrices
declared in the namespace ``tensor2tools'' (see the document
for details on this namespace).
\begin{Verbatim}
#include "zkcm_qc.hpp"

int main (int argc, char *argv[])
{
  //Use the 256-bit float for internal computation.
  zkcm_set_default_prec(256);
  //Num. of digits for each output is set to 8.
  zkcm_set_output_nd(8);

  //First, we make an MPS representing |000>.
  mps M(3);
  std::cout << "The inital state is " << std::endl;
  //Print the reduced density operator of the block
  //of qubits from 0 to 2, namely, 0,1,2, using the
  //binary number representation for basis vectors.
  tensor2tools::showb(M.RDO_block(0,2));

  std::cout << "Now we apply H to the 0th qubit."
            << std::endl;
  M.applyU(tensor2tools::Hadamard, 0);

  std::cout<< "Now we apply CNOT to the qubits 0 and 2."
           << std::endl;
  M.applyU(tensor2tools::CNOT, 0, 2);

  //The array is used to specify qubits to compute
  //a reduced density matrix. It should be terminated
  //by the constant mps::TA.
  int array[] = {0, 2, mps::TA};
  std::cout << "At this point, the reduced density\
 matrix of the qubits 0 and 2 is " << std::endl;
  tensor2tools::showb(M.RDO(array));

  return 0;
}
\end{Verbatim}

The output of the program is as follows.
\begin{Verbatim}
[user@localhost samples]$ ./qc_simple_example
The inital state is
1.0000000e+00|000><000|
Now we apply H to the 0th qubit.
Now we apply CNOT to the qubits 0 and 2.
At this point, the reduced density matrix of \
the qubits 0 and 2 is
5.0000000e-01|00><00|+5.0000000e-01|00><11|\
+5.0000000e-01|11><00|+5.0000000e-01|11><11|
\end{Verbatim}

\section{Summary}\label{secSummary}
In this report, a C++ library ZKCM for multiprecision complex-number
matrix calculation has been introduced. It reduces the cost of writing
elaborate programs especially in case a small deviation is of main
concern, which is often the case for time-dependent physical models.
An extension library ZKCM\_QC has also been introduced, which is a
library for an MPS simulation of quantum circuits. It is designed to
enable an intuitive coding manner to simulate quantum circuits.

\subsection*{Software information}
\noindent ZKCM and ZKCM\_QC libraries are open-source C++ libraries.
The files and documents can be downloaded from the URL
shown as Ref.\ \cite{ZKCM}. ZKCM version 0.0.9 and ZKCM\_QC version
0.0.1 on the Fedora 15 64-bit operating system with GMP version 4.3.2
and MPFR version 3.0.0 have been used for this report. 

\subsection*{Acknowledgment}
\noindent A. S. is supported by the ``Open Research Center''
Project for Private Universities: matching fund subsidy from MEXT.

\renewcommand{\refname}{References}


\begin{thebibliography}{99}
\bibitem{FMZM} D. M. Smith, Multiple Precision Complex Arithmetic and
	Functions, Trans. Math. Software {\bf 24}, 359-367 (1998),
\mbox{http://myweb.lmu.edu/dmsmith/FMLIB.html}
\bibitem{CLN} Maintained by B. Haible and R. B. Kreckel, CLN - Class
	Library for Numbers, \mbox{http://www.ginac.de/CLN/}
\bibitem{Exflib} H. Fujiwara, exflib - extend precision floating-point
	arithmetic library,\\
	\mbox{http://www-an.acs.i.kyoto-u.ac.jp/$\sim$fujiwara/exflib/}
\bibitem{MPACK} M. Nakata, The MPACK (MBLAS/MLAPACK); a multiple
	precision arithmetic version of BLAS and LAPACK, \mbox{http://mplapack.sourceforge.net/}
\bibitem{ZKCM} A. SaiToh, ZKCM and ZKCM\_QC,\\
	\mbox{http://zkcm.sourceforge.net/}
\bibitem{GMP} The GNU Multiple Precision Arithmetic Library,\\ \mbox{http://gmplib.org/}
\bibitem{MPFR} The GNU MPFR Library, \mbox{http://www.mpfr.org/}
\bibitem{Gruska}J. Gruska, \textit{Quantum Computing} (McGraw-Hill, London, 1999).
\bibitem{NC2000}M. A. Nielsen and I. L. Chuang,
\textit{Quantum Computation and Quantum Information}
(Cambridge University Press, Cambridge, England, 2000).
\bibitem{WH93}S. R. White,
Density-matrix algorithms for quantum renormalization groups,
Phys.\ Rev.\ B \textbf{48}, 10345 (1993).
\bibitem{V03}G. Vidal,
Efficient classical simulation of slightly entangled quantum computations,
Phys.\ Rev.\ Lett. {\bf 91}, 147902 (2003).
\bibitem{ALPS}B. Bauer {\em et al.}, The ALPS project release 2.0: open source
software for strongly correlated systems, J.\ Stat.\ Mech. {\bf 2011}(05), P05001 (2011),
\mbox{http://alps.comp-phys.org}
\bibitem{VMH03}G. F. Viamontes, I. L. Markov, and J. P. Hayes,
Improving Gate-Level Simulation of Quantum Circuits,
Quantum Inf. Process. {\bf 2}(5), 347, (2003).
\bibitem{AG04}S. Aaronson and D. Gottesman,
Improved simulation of stabilizer circuits,
Phys.\ Rev.\ A {\bf 70}, 052328 (2004).
\bibitem{KW04}A. Kawaguchi, K. Shimizu, Y. Tokura, and N. Imoto,
Classical simulation of quantum algorithms using the tensor product representation,
e-print arXiv:quant-ph/0411205.
\bibitem{S06}A. SaiToh and M. Kitagawa,
Matrix-product-state simulation of an extended Br\"uschweiler bulk-ensemble database
 search,
Phys. Rev. A {\bf 73}, 062332 (2006).
\bibitem{GAMMA} S. A. Smith, T. O. Levante, B. H. Meier, and
	R. R. Ernst, Computer Simulations in Magnetic Resonance. An Object Oriented Programming Approach,
J.\ Magn.\ Reson., {\bf 106a}, 75-105, (1994), \mbox{http://gamma.ethz.ch/}
\end{thebibliography}
\end{document}